\documentclass[conference]{IEEEtran}
\ifCLASSINFOpdf
\else
\fi

\usepackage{epsfig}
\graphicspath{{Figures/}}

\begin{document}
%
\title{Contribution of Different Handwriting Modalities to Differential Diagnosis of Parkinson's Disease }

\author{\IEEEauthorblockN{Peter Drot\'ar, Ji\v{r}\'i Mekyska, Zden\v{e}k Sm\'ekal }
\IEEEauthorblockA{Faculty of Electrical Engineering and Communications\\
Brno University of Technology\\
Brno, Czech Republic\\
Email: drotarp@feec.vutbr.cz}
\and
\IEEEauthorblockN{Irena Rektorov\'a, Lucia Masarov\'a}
\IEEEauthorblockA{First Department of Neurology\\
 CEITEC MU\\
Masaryk University\\
Brno, Czech Republic}
\and
\IEEEauthorblockN{Marcos Faundez-Zanuy}
\IEEEauthorblockA{Tecnocampus\\
Escola Universitaria Politecnica \\
de Mataro\\
Mataro, Spain}}


%


\maketitle

\begin{abstract}
In this paper, we evaluate the contribution of different handwriting modalities to the diagnosis of Parkinson's disease. We analyse on-surface movement, in-air movement and pressure exerted on the tablet surface. Especially in-air movement and pressure-based features have been rarely taken into account in previous studies. We show that pressure and in-air movement also possess information that is relevant for the diagnosis of Parkinson's Disease (PD) from handwriting. In addition to the conventional kinematic and spatio-temporal features, we present a group of the novel features based on entropy and empirical mode decomposition of the handwriting signal. The presented results indicate that handwriting can be used as biomarker for PD providing classification performance  around $89 \%$  area under the ROC curve (AUC) for PD classification.
\end{abstract}


%
\IEEEpeerreviewmaketitle

\section{Introduction}
According to the recent estimates, more than seven million people are affected by Parkinson's Disease (PD) worldwide~\cite{deLau_epidemiology}. The high number of affected people makes PD the second most common neurodegenerative disorder. Moreover, with the aging population, it is expected that the prevalence rates will further increase and impose social and economic burden for healthcare. Despite intensive research effort, the causes of PD are still not known and the reliable easily applicable diagnostic test is not available yet~\cite{deLau_epidemiology}. 

A diagnosis of PD is currently based mainly on clinical symptoms such as bradykinesia, rigidity, tremor or postural imbalance. Several alternative solutions and decision support systems have been proposed to improve diagnosis of PD. The neuroimaging methods show significant potential but require expensive equipment~\cite{wu}. Other approaches include attempts to detect PD from breath \cite{tisch} or voice~\cite{tsanas_lsvt}, \cite{rusz_pdSpeech}, \cite{mekyska_pdSpeech}. Especially speech processing for diagnosis of PD gained significant attention and offered very promising results. Recent studies indicate that also handwriting can be with advance used for differential diagnosis of PD~\cite{rosenblum_parkinson},~\cite{drotar_ieeet},~\cite{smits_hw}. This is due to the alterations in parkinsonian handwriting represented by micrographia and PD dysgraphia \cite{letanneux_pd}. Micrographia was reported for the first time by Pick~\cite{pick} as an abnormal reduction in handwriting size associated with PD. On the other hand, PD dysgraphia is a new term proposed by Letanneaux et al.~\cite{letanneux_pd} that "\textit{encompasses the whole spectrum of disorders that affect the writing of PD patients}" including tremor, rigidity, bradykinesia, akinesia, freezing of the upper limb etc.

Micrographia was studied in \cite{Gemmert_pd} or \cite{broderick_hypometria}, however analyzing only micrographia alone is not enough, since micrographia occurs only in 30 \% to 60 \% of patients with PD \cite{kim_micrographia},\cite{contreras_micrographia}. This was motivation for several authors to investigate also kinematic aspects of movement including speed, acceleration or stroke duration \cite{Teulings_acceler}, \cite{unlu_hw}. Even though these studies provided the significant insight into the handwriting in PD, they did not assess tremor. Recently, two studies have been published that provide quantitative measures to assess multiple motor symptoms of the PD handwriting allowing for the clinically acceptable differential diagnosis of PD \cite{smits_hw},\cite{drotar_ieeet}. The current technologies allow to exploit new modalities, such as in-air movement and pressure exerted on the surface, not only conventional handwriting trace on surface~\cite{nogueras_information}. Some initial studies employing the in-air movement \cite{drotar_bibe}, \cite{rosenblum_parkinson} or pressure \cite{rosenblum_parkinson} indicate that also these modalities can be useful for diagnosis of PD.  

In this study, we employ an approach presented in our previous work \cite{drotar_ieeet} and use advanced handwriting markers based on entropy, energy and intrinsic measures of handwriting. Here, we apply these measures also to in-air movement and pressure to exploit full potential of the handwriting for classification of PD. To achieve this goal we make use of our Parkinsonian Handwriting database (PaHaw) which consists of seven different handwriting tasks. The task template contains the exercises found in similar studies that are used to assess PD. In addition to conventional handwriting tasks we added the new tasks: simple words and one sentence.  The orthography of these task is intentionally simple to minimize cognitive effort during the writing task.

The paper is organized as follows. After the introductory section, the description of used database is given. Next, methodology of extracting the features from handwriting signal is given, followed by a brief overview of used classifier. Finally, the numerical results and conclusions are provided.

\section{Data}

The Parkinsons handwriting dataset consists of multiple handwriting samples from  37 parkinsonian patients (19 men/18 women) and 38 gender and age matched controls (20 men/18 women). Mean age was $69.3\pm 10.9$ for parkinsonian patients and $62.4\pm 11.3$ for control subjects, respectively. All subjects were right-handed, had completed at least 10 years of education, and reported Czech as their first language.  For parkinsonian patients the mean value of Unified Parkinson's Disease Rating Scale-Part~V was $2.27\pm 0.84$ and all patients completed the tasks under medication L-DOPA. A more detailed description of dataset is shown in \cite{drotar_ieeet}.

\begin{figure} 
	\centering
    \includegraphics[width= 0.9\linewidth]{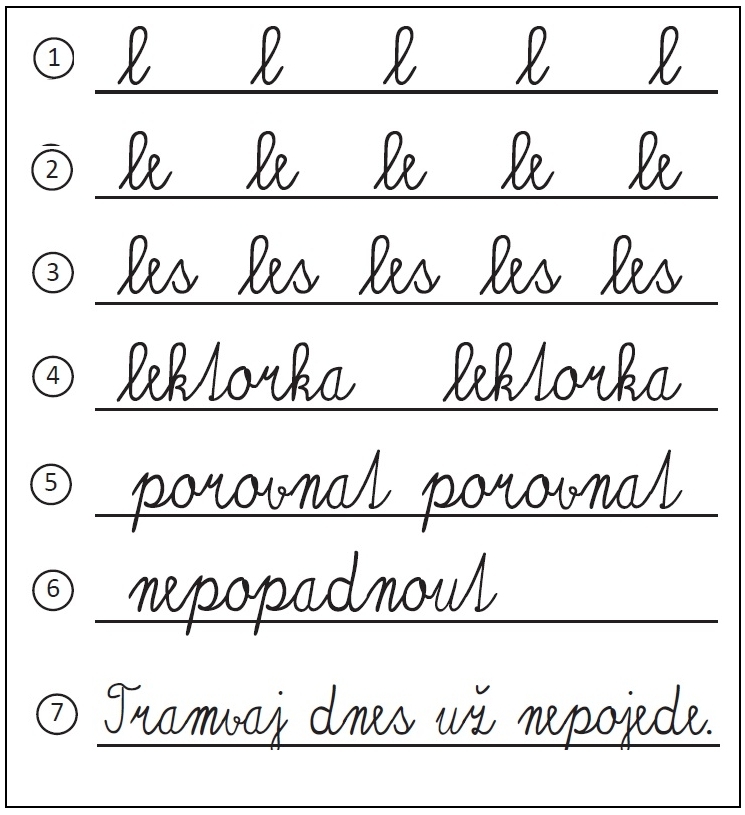} 
    \caption{Filled task sheet used as template for handwriting tasks. }
    \label{fig_template}
\end{figure}

Each subject was asked to complete handwriting task according to the prepared template. The template consists of 7 tasks that were a part of a more detailed test battery. The tasks were designed to have simple orthography and all but last one can be written as one long stroke. The filled task sheet is depicted in~Fig.\ref{fig_template}. The template was shown to the patients and they were free to write the words without no need to follow the exact pattern.

\section{Feature extraction}

Signals were acquired using Wacom Intuos 4M pen tablet. The tablet itself does not provide visual feedback, therefore, it was overlaid with the paper so the pen can be held in a normal fashion and allows full visual feedback during writing. The inked pen is commercially available from the tablet producer, it allows a user to draw on the paper and at the same time the signal is captured by the tablet. 

\begin{figure*} 
	\centering
    \includegraphics[width= 1\linewidth]{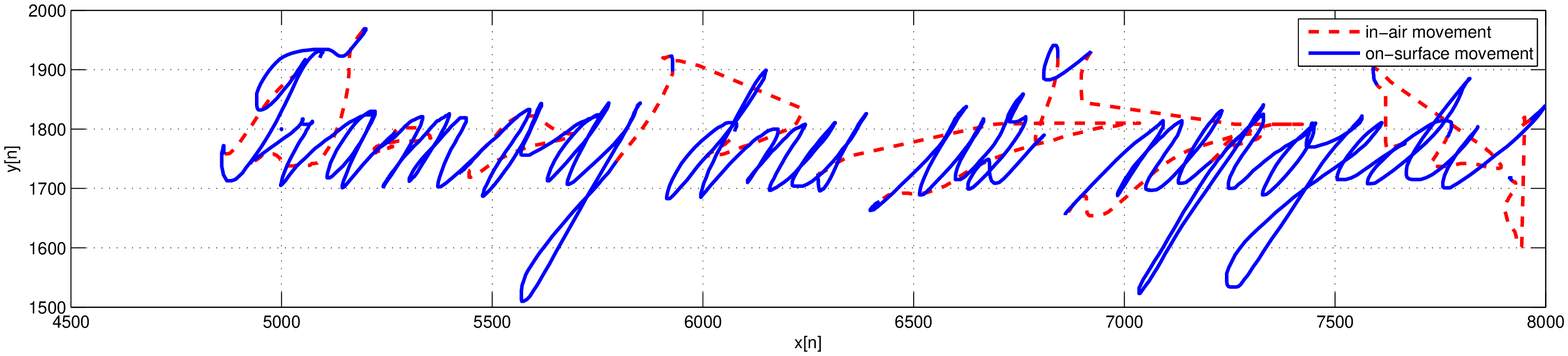} 
    \caption{Handwriting sample from seventh task of the template. In-air and on-surface movement. }
    \label{fig_task_7}
\end{figure*}

\begin{figure*} 
	\centering
    \includegraphics[width= 1\linewidth]{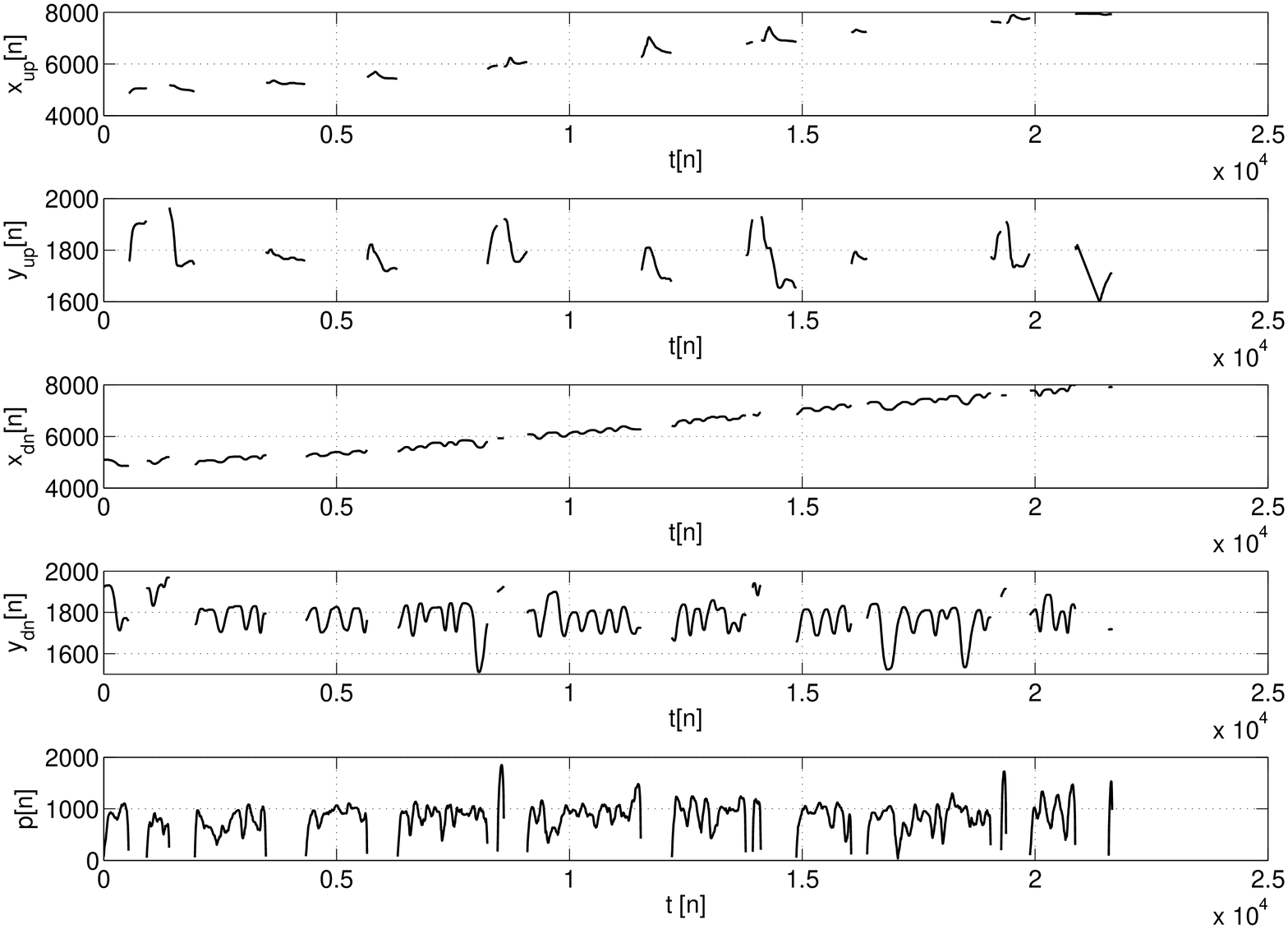} 
    \caption{Recorded signals during the execution of the seventh task. The figure displays x and y coordinates of in-air (up) and on-surface (dn) movement as a function of time. Additionally bottom figure illustrates pressure exerted on tablet surface as an function of time.  }
    \label{fig_signals}
\end{figure*}

The tablet is able to capture several signals related to the handwriting. These include position of the pen tip in the form of x and y coordinate ($x[n]$, $y[n]$), binary variable ($b[n]$), being 0 for pen-up state (in-air movement) and 1 for pen-down state (on-surface movement). This means that the tablet is able to capture $x[n]$ and $y[n]$ also in case when pen is not touching the tablet surface while moving in the air. Additionally, pressure exerted on the tablet surface during writing ($p[n]$) and time stamp ($t[n]$) are recorded. An example of handwriting sample from the 7th task is depicted in Fig. \ref{fig_task_7}. The figure shows the example of recorded on-surface (blue solid line) and in-air (red dashed line) movement. Even though it illustrates how the actual handwriting sample is realized it does not provide the insight into the shape of the signals that were used to extract handwriting features. Signals corresponding to the sample depicted in Fig. \ref{fig_task_7} can be seen in Fig. \ref{fig_signals}. Fig. \ref{fig_signals} shows pressure and coordinates of in-air/on-surface movement as a function of time. 


\subsection{Spatio temporal and kinematic features}

Using Cartesian coordinates ($x[n]$,$y[n]$) and time stamp it is possible to determine several kinematic and spatio-temporal features. These include handwriting velocities and derived measures such as acceleration and jerk: speed, a number of changes in velocity(NCV)/acceleration(NCA), relative NCV/NCA, stroke speed, velocity, acceleration, jerk, horizontal velocity/acceleration/jerk, vertical velocity/acceleration/jerk. Considered spatio-temporal features are stroke height/width, writing duration, writing length and in-air to on-surface ratio. The definition of these features is provided in \cite{drotar_bibe}.

\subsection{Pressure features}

To make use of recorded pressure signal $p[n]$ several pressure features were extracted. Similarly to kinematic features, we computed rate at which pressure changes with respect to time, a number of changes in velocity pressure (NCP) and relative NCP. Relative NCP is NCP normalized by the whole writing length. Additionally, six correlation coefficient were introduced: correlation between pressure and (horizontal/vertical) velocity/acceleration. Fig.~\ref{fig_signals} indicates that typical pressure stroke starts with rising edge, continues with slowly varying main part and ends with falling edge. We derived pressure features for each part of the stroke (edge, main part of signal and falling edge) separately. The boundary between edges and main part is given by median of signal pressure. Finally, a range of rising and falling edge in terms of pressure and time was included in the analysis.

\subsection{Nonstandard handwriting features}

In order to uncover also hidden complexities of handwriting we also employ nonstandard features proposed in \cite{drotar_ieeet}. Out of these, entropy based features have potential to capture randomness of the movement during handwriting. We calculated the Shannon entropy $H_S$ and R\'enyi entropy of the second $H_{R,2}$ and the third $H_{R,3}$ order~\cite{cover_book}.	  

Similarly, the energy based features express amount of noise in handwriting in relation to useful handwriting signal. Note that what we refer to as noise is not unwanted interference from an external source, but the irregular movement resulting from muscle contractions and irregularities. Therefore, we computed estimated noise variance $\mathcal{N}(s[n])$ of the signal and energy of the signal $\Theta (s[n])$. The operator $\Theta$ represents conventional energy operator $CE(s[n]) = 1/N\sum\limits_{n=0}^{N-1} s[n]^2$ or Teager-Kaiser energy operator $TKE_r(s[n]) = 1/N\sum\limits_{n=0}^{N-r-1} s[n]^2 - s[n+r]\cdot s[n-r]$. Signal to noise ratios are given by $SNR_{\Theta} = \Theta(s[n])/ \mathcal{N}(s[n])$.
The signal $s[n]$ is the signal under evaluation i.e. $x[n]$, $y[n]$ (obtained from both on-surface and in-air movement) or $p[n]$.

To obtain intrinsic features we apply empirical mode decomposition (EMD) to decompose signal into its intrinsic mode functions (IMF). EMD is an intuitive data-dependent de-composition of a time series that allows decomposition of non-linear and non-stationary data into IMFs. EMD is conducted by the iterative extraction based on the local representation of the signal as the sum of a local oscillating component and a local trend \cite{Huang_emd}. Given a time series $s[n]$, combining all the IMFs gives the original signal $s[n] = \sum_{j=1}^N IMF_j[n]+r_N[n]$. 

We obtained intrinsic entropies and energies by applying the above mentioned methods on first and second intrinsic function. Similarly, intrinsic signal to noise rations were derived as the ratio of energy of first two intrinsic functions to energy of the rest of intrinsic functions. 
 
\section{Machine Learning}

Six ”basic” statistical functionals (mean, median, standard deviation, 1st percentile, 99th percentile, 99th percentile - 1st percentile) were computed from the extracted features. This produces more than 5000 features. The features were normalized before classification on a per-feature basis to have zero mean and a standard deviation of one. As a preprocessing step we applied Mann-Whitney U test for significant differences identification to remove features that did not show statistical significance to class label. Application of Mann-Whitney U test reduced the number of features to less than 700. The distribution of features for different tasks and modalities is depicted in Tab. \ref{tab_features}.

\begin{table*}
\centering
\caption{Number of extracted features and number of features that passed the Mann-Whitney U test}
\label{tab_features}       
\begin{tabular}{cccc}
\hline\noalign{\smallskip}
 & on-surface (task 1 - task 7) &  in-air (task 1 - task 7) & pressure (task 1 - task 7)\\

all & 268 per task & 290 per task & 188 per task  \\
afer Mann-Whitney U test & 11, 61, 12, 10, 10, 35, 138 &  58, 81, 2, 7, 8, 0, 80 & 32, 39, 11, 10, 24,  29, 34 \\
\noalign{\smallskip}\hline
\end{tabular}
\end{table*}

Support Vector Machine (SVM) was used as a classifier to predict class labels. The underlying idea of SVM classifiers is to calculate a maximal margin hyperplane separating two classes of the data. To learn non-linearly separable functions, the data are implicitly mapped to a higher dimensional space by means of a kernel function. The new samples are classified according to the side of the hyperplane they belong to. We used Radial Basis Function (RBF) kernel~\cite{vapnik}. The RBF kernel is defined as
\begin{equation}
K(x,x_i)= \mathrm{e}^{\frac{- \|x-x_i\|^2 }{2\gamma^2}}
\end{equation} 
where $\gamma$ controls the width of RBF function.

The parameters kernel gamma $\gamma$ and penalty parameter $C$ were optimized using grid search of possible values.  Specifically, we searched over the grid $(C, \gamma)$ defined by the product of the sets $C = [2^{-10}, 2^{-9}, \ldots, 2^{6}, 2^{7}]$, $\gamma = [2^{-7}, 2^{-6}, \ldots, 2^{6}, 2^{7}]$. We used scikit-learn implementation of SVM \cite{scikit-learn}.

\section{Numerical Results and Discussion}

The prediction performance was evaluated in terms of the area under ROC curve (AUC). In order to obtain prediction potential of each modality we assess the AUC performance for every modality (in-air, on-surface and pressure) individually. Additionally, prediction performance of seven different handwriting tasks was also considered separately. The numerical results achieved by SVM classifier with 10 fold cross validation are provided in Table \ref{tab_results}.

As it can be seen from Table \ref{tab_results}, the highest $AUC=89.09 \%$ was achieved when the features extracted from on surface movement were used. This provides significantly higher AUC than other two modalities. The promising results are also yielded from the pressure features that have not been used for purpose of PD classification before, giving rise to more than $83 \%$ prediction accuracy. Within this scenario in-air movement does not appear to be significantly contributing to differential diagnosis of PD from handwriting. However, we should note that entropy, energy and intrinsic features were originally designed for on-surface movement and as such may not explore full potential of in-air movement.

When comparing the contribution of a different handwriting task to classification of PD, it is clear that most of the prediction performance comes from the seventh task. This is true especially for the on-surface and in-air modality.  In case of pressure, different handwriting tasks contribute more equally. The seventh task (\textit{Tramvaj dnes u\v{z} nepojede)} is the longest task and in contrast to other tasks this task cannot be written as one long stroke. Writing longer sentence probably requires higher cognitive effort and escalates effect of disease on handwriting. These results are in agreement with the previous findings where the last task also appears to be the most predictive one~\cite{drotar_ehb}.  

We tried to identify a smaller subset of the features that provide the strongest discriminative power. However, reducing number of the features resulted in decline in classification performance. 

\begin{table}[t!]
\caption{AUC of PD classification based on different modalities (in-air, on-surface and pressure).}
\label{tab_results}       
\centering
\begin{tabular}{cccc}
\hline\noalign{\smallskip}
 task/modality & on-surface & in-air & pressure \\ \hline \hline
1			   &	72.39	& 67.58  & 72.5		\\	 
2              & 	70.16	& 66.75  & 76 		\\
3              &	70.86	& 66.75  & 72.16		\\	
4              &	66.08	& 65.25  & 64.25	\\
5              &	62.75	& 67.33  & 69.66	\\
6              & 	65.66	&  -     & 71.66	\\
7              & 	83.83	& 73	 & 72.58	\\ \hline
all			   &	89.09	& 74.16  & 83.83	\\
\noalign{\smallskip}\hline
\end{tabular}
\end{table}

\section{Conclusion}

We have evaluated a prediction performance of the different handwriting modalities for differential classification of PD. It was shown that not only on-surface movement, but also pressure and in-air movement contribute to the classification and can be with advantage used for diagnosis of PD from handwriting. By using 7 different handwriting tasks, standard kinematic and also novel intrinsic and entropy features we showed that handwriting is a promising tool for diagnosis of PD achieving almost $90\%$ prediction performance.


\section*{Acknowledgment}

This work was supported by the projects COSTIC1206,NT13499 (Speech, its impairment and cognitive
performance in Parkinson’s disease), CZ.1.07/2.3.00/20.0094,
project CEITEC, Central European Institute of Technology:
(CZ.1.05/1.1.00/02.0068) from the European Regional Devel-
opment Fund and by FEDER and Ministerio de Economa
y Competitividad TEC2012-38630-C04-03. The described
research was performed in laboratories supported by the
SIX project; the registration number CZ.1.05/2.1.00/03.0072,
the operational program Research and Development for
Innovation. Peter Drot\'ar was supported by the project
CZ.1.07/2.3.00/30.0005 of Brno University of Technology.



%
\bibliographystyle{IEEEtran.bst}
\bibliography{../_BIBLIOGRAPHY_/pd_bib}

\end{document}